\documentclass{iaus}
\usepackage{graphicx}
\usepackage{natbib}

\title[M31 Luminosity Functions] 
{Cumulative luminosity functions of the X-ray point source population
  in M31}

\author[L. Shaw Greening et al]   
{L. Shaw Greening,$^1$ %
 \,C. Tonkin,$^1$  R. Barnard,$^1$  U. Kolb$^1$ \break \and J.P. Osborne$^2$}

\affiliation{$^1$Open University,
  Walton Hall, Milton Keynes, MK7 6AA, UK \break email: L.Shaw-Greening@open.ac.uk\\[\affilskip]
$^2$University of Leicester, University Road, Leicester, LE1 7RH, UK}

\pubyear{2005}
\volume{230}  
\pagerange{119--126}
\date{?? and in revised form ??}
\jname{Proceedings Title IAU Symposium}
\editors{J.A. Meurs \& Giuseppina Fabbiano}
\begin{document}

\maketitle
\vspace{-0.25in}
\section{Introduction}
We present preliminary results from a detailed analysis of the X-ray point
sources in the XMM-Newton survey of M31 \citep[e.g.][]{Barnard05,
  Pietsch05}.  These sources are expected to be mostly X-ray binaries.   

\vspace{-.25in}
\section{Source Detection}
\begin{figure}
\begin{center}
\includegraphics[scale=0.6]{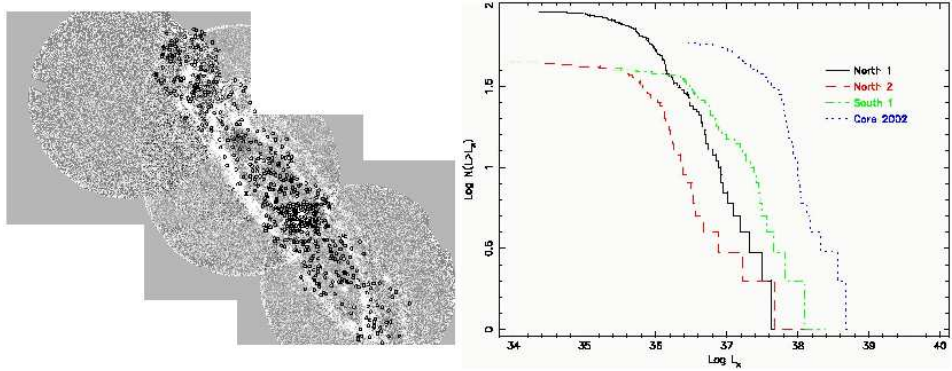}
\end{center}
\caption{The left hand panel shows the X-ray sources detected,
  superimposed on a mosaic of GALEX FUV observations of M31.  The
  right hand panel shows the luminosity functions of each of the fields
  we have studied.}\label{LF}
\end{figure}

The left hand panel of Fig.~\ref{LF} shows a mosaic of GALEX FUV images of
M31 \citep{Thilker05} overlaid by the point sources detected by
edetect\_chain in SAS 6.0.0.  Only sources which were on both the PN and
MOS chips were analysed.
The minimum detection likelihood was set to 10 and detections
that overlapped at a radius of 20'' were excluded. Lightcurves, energy
spectra and power density spectra were then extracted for each region
and analysed to determine source properties. Energy spectra were fit with
power law, bremsstrahlung, black body and neutron star atmosphere
models in xspec 11.3.1 with free parameters to determine the flux of
the source.  The unabsorbed luminosities were then calculated assuming
a distance to M31 of 760 kpc \citep{Bergh00}. 

\vspace{-0.25in}
\section{Results}
In the right hand panel of Fig.~\ref{LF}, we present  the first
cumulative luminosity functions of the XMM-Newton M31 fields
to be created using individual fits for each source spectrum to obtain
the unabsorbed luminosity for each source.
The North 1 (solid) and 2 (dashed) field luminosity functions are
very similar in shape despite having very different numbers of
sources.  The Core and South 1 luminosity functions have different
shapes because these fields have not yet been fully surveyed down to
the same luminosity limits as the North fields.  The
final luminosity limit for each field will also depend on the
good-time interval of each observation.  We will investigate the
significance of these luminosity functions once the survey is complete.

We have also created a luminosity function of the North 1 and 2 fields
combined in order to compare it to the luminosity function of these
fields reported by
\citet{Trud02}.  Unlike in our study, these authors used individual
spectral fits only for the brightest sources ($L_X>5
\times 10^{36}$ erg s$^{-1}$); the faint sources had their luminosities
estimated assuming an absorbed power law model with fixed power law
index. Above L$_X$=10$^{36}$ erg s$^{-1}$ our luminosity function can
be approximated by a power law $N(L>L_X) \thicksim L_X^{-\alpha}$ with
slope $\alpha$=0.87, inconsistent with $\alpha$=1.3 found by
\citet{Trud02}.  This is probably a consequence of the fixed spectral
model assumed by these authors.

\citet{Kong02} used Chandra to study the core region of M31.  They
extracted energy spectra from the longest (8.8ks) observation and
carried out spectral fitting only on the brightest 20 of the 204
sources they found.  All sources were well fit by both absorbed power
law and Raymond-Smith models.  For the other 184 sources, count rates
were converted to luminosities assuming an absorbed power law model.
They obtained a luminosity function described by $\alpha$=1.44 above a
break at 1.77 $\times 10^{37}$ erg
s$^{-1}$ for the whole core exposure. This is comparable with our
$\alpha$=1.96 for $37.7<$log(L$_X$ erg s$^{-1})<38.2$ in the Core
luminosity function.

\citet{Grimm02} studied the Milky Way and created
luminosity functions for the high and low mass binary populations
separately.  Their cumulative luminosity functions had $\alpha$=0.64
for HMXBs and $\alpha$=0.26 for LMXBs.  Neither of
these slopes is consistent with our luminosity functions of M31 but the binary
population cannot be easily separated in M31 and both types of binary
are present.

\vspace{-0.25in}
\section{Future Work}
We have so far studied 225 sources over 4 fields and our survey is
continuing by looking at the unstudied sources we have detected in the
2 outermost disc fields, the 130 fainter core sources and the 51 fainter
sources in the South 1 field. X-ray binaries will be identified by
their energy spectrum and 
power density spectrum as outlined in \citet{Barnard04,Barnard05}.  We
will also continue to create luminosity functions for each field to
allow us to compare populations. All the luminosity functions we
create  will also
need to be corrected for foreground and background sources
\citep{Giacconi01, Hasinger01} and for incompleteness (see \citet*{KF04})

\end{document}